# The Epistemic Virtues of the Virtuous Theorist: On Albert Einstein and His Autobiography[1]

Jeroen van Dongen

*Institute for Theoretical Physics Amsterdam*
*Vossius Center for the History of Humanities and Sciences*
*University of Amsterdam, Amsterdam, The Netherlands*

**Abstract**

Albert Einstein's practice in physics and his philosophical positions gradually reoriented themselves from more empiricist towards rationalist viewpoints. This change accompanied his turn towards unified field theory and different presentations of himself, eventually leading to his highly programmatic *Autobiographical Notes* in 1949. Einstein enlisted his own history and professional stature to mold an ideal of a theoretical physicist who represented particular epistemic virtues and moral qualities. These in turn reflected the theoretical ideas of his strongly mathematical unification program and professed Spinozist beliefs.

**Introduction**

In the early 19th century, the experimental physicist Michael Faraday perfected his notebook recordkeeping such that the data would enter them entirely without regard to what his original hypotheses might have been. His way of working sharply contrasts with Arthur Worthington's, who in the earlier years of his career saw the need to generalize and brush over asymmetries in his visual studies of liquid drops: such asymmetries were deemed irrelevant for capturing the latter's essence. Faraday and Worthington represent two ideal types, two 'personae' that figured prominently in the practice of nineteenth century science. The novel and self-denying scientist aspiring to the objective representation of nature represented the world differently than the intuitively working scholar, who wished to point out the true essence of a natural phenomenon: the first ideally presented his observations unfiltered and directly, without intervention; the second would at times, e.g., see the need to smooth out the irregular or asymmetric.

This is of course one of the key observations of Lorraine Daston and Peter Galison's study *Objectivity* (2007, 243-246). Clearly, the dual roles of the hypothesizing and observing scientist changed due to the rise of the epistemic virtue of objectivity in the course of the nineteenth century. The examples of Faraday and Worthington may be taken to also illustrate another key aspect of the role of epistemic virtues, however: they document how the nature and content of the *knowledge produced* changed as a different epistemic virtue, embodied by a different scholarly persona, rose to prominence.

---

[1] Published as: Jeroen van Dongen, "The epistemic virtues of the virtuous theorist: On Albert Einstein and his autobiography", pp. 63-77 in J. van Dongen and H. Paul (eds.), *Epistemic Virtues in the Sciences and the Humanities* (Boston Studies in the Philosophy and History of Science, vol. 321), Cham: Springer, 2017.

This connection between knowledge and the roles of epistemic virtues and personae[2] is often overlooked. Nevertheless, it may reveal us hitherto unseen relations in knowledge production. To illustrate the point, we will study the example of Albert Einstein, who used his own biography and related ideal versions of the 'theoretical physicist' to argue for the virtuousness of his 'unified field theory' physics.

What is a 'scientific persona' and how may this concept relate to Einstein and his autobiography? Literature on 'personae' places the concept between individual biography and social institution: 'personae' (for example, the 'scientist' vs. the 'natural philosopher' vs. the 'poet') mark a cultural identity that is shared and serves as example among a collective body of scholars (Daston and Sibum, 2003). A particular persona is reflective of a set of values, attitudes and scientific practices; it is both negotiated, internalized and projected among the practitioners of a discipline. It represents considerable agency and influence as disciplinary icon. In Einstein's time, theoretical physics was a fairly new sub-discipline,[3] coming into its own on the heels of the relativity successes: the persona of the 'theoretical physicist' was being molded while that of a 'physicist' *per se* was being reshaped. As the resistance to relativity exemplifies (see, e.g., Wazeck 2009; van Dongen 2010b), Einstein's science, biography and public stature were central in the realignment of what an ideal physicist and 'theorist' should be: abstractions of Einstein's person were projected on these personae while they were being (re-)created.[4]

As Einstein cared deeply about the future direction of physics and the plight of his later endeavours, he inadvertently but deliberately engaged his own history to argue for particular scientific choices and certain notions of what a proper theorist was and did. In various autobiographical accounts, and particularly in his 1949 *Autobiographical Notes* Einstein presented certain choices for theoretical virtues and methods as linked to certain epistemically virtuous attitudes. Furthermore, these choices were being promoted as representing both the epistemically and the *morally* right choice. On this point, one is reminded of Thomas Kuhn's essay on theory choice: in Kuhn's analysis, the weighing of various virtues such as 'simplicity' or 'accuracy' in theory choice is just *like* the weighing of values in moral judgment (Kuhn 1977). The example of Einstein will suggest that the relationship between these kinds of judgments need not only be seen as that of a mere analogy, but as at times direct and concrete.

This paper will then do two things: it will first show how indeed the promotion of particular epistemic virtues can be linked with concrete choices made in the practice of science, and thus how they affect its result: the choice of theory. We will study this relation in Einstein's motivations for his unified field theories. Secondly, we will see in Einstein's example how the distinctions between the moral and the cognitive may dissolve in the role of these virtues, and how in his case they were linked to the conviction that 'true' scientific pursuit is preconditioned on a religious gaze.

---

[2] See in this regard also Tai and van Dongen 2016; Tai 2017.

[3] For the sub-discipline's history, see Jungnickel and McCormmach 1986.

[4] See for example how young Werner Heisenberg related to Einstein and the latter's recollections in the mid-1920s: Heisenberg [1974] 1989, in particular 113-114.

Strikingly, the relation between theory choice, theoretical virtues and epistemic virtues is laid bare when they meet in the ideal Einstein portrayed of himself later in life. What virtues can we identify and what was their role in his practice of science? Given Einstein's actual history and agenda, how can we understand the autobiographical accounts that he produced? Clearly, addressing these questions will not only help in understanding the relation between knowledge and the persons making it, but also aid in forming a comprehensive biographical and literary coherent understanding of Einstein. We begin by looking at Einstein's turn to unified field theory[5] and how this fits into his broader development as a scholar.

**Einstein's *Werdegang* from empiricist to rationalist**

In 1905, his 'miracle year', Einstein famously proposed the first version of the theory of relativity, as it would later be known. It was essentially a reformulation of the electrodynamics of Hendrik Antoon Lorentz, but a reformulation in which experience stood at the epistemic core. Central was a new interpretation of the time parameter, which was motivated by a Machian sense-datum type intuition: "we should be aware that all our pronouncements in which time plays a role are always pronouncements about simultaneous events. For example, if I say that 'that train arrived here at 7 o'clock', I mean to say, more or less, that the arrival of the train and the pointing of the small hand of my clock at 7 are simultaneous affairs" (Einstein 1905, 892-893). Clearly, the author of this passage stands directly in the world of experience and looks around himself, from whence he constructs his concepts.

Later in life, in 1954, Einstein opined in a letter to a colleague, Louis de Broglie, that he had "long been convinced that one shall not be able to find [the right description of the quantum] in a constructive way from the known empirical relations between physical things, because the required mental leap would exceed human powers."[6] In this statement, experience and observation are put at a distance from the creative scientific process. The move reflects Einstein's then decades long involvement with unified field theory research. These theories— Einstein's attempts at an alternative theory for the quantum—were foremost creative efforts based on mathematics and hardly made any connection to experience.

Einstein's early work, however, often offered direct experimental hypotheses, as in the case of his light quantum proposal of 1905. Revealingly, in 1925 Einstein wrote to his friend and fellow theoretical physicist Paul Ehrenfest on that very same subject that he "no longer think[s] about experiments on the boundary of waves and particles" as "inductive means will never get you to a sensible theory."[7] Einstein's change in attitude regarding the epistemic role of experience is also reflected in the difference in appreciation he awarded 'facts' (i.e. experientially validated facts about the world) in the recollection of his own struggle to attain his greatest achievement, the formulation of the general theory of relativity, completed in 1915. In 1918, he held that that struggle had taught him that trustworthy theories needed "to be built on generalizable facts."[8] In the 1949 autobiography, however, Einstein expressed that general relativity had taught him something entirely different: that there was "no way" to formulate a successful theory from a "collection of facts" (Einstein [1949], 89). Experience could only play a role at the end of the creative process, to check whether its product was valid or not.

---

[5] For a more extensive account of this aspect, see van Dongen 2010a.

[6] Einstein to Louis de Broglie, 8 February 1954, as quoted in van Dongen 2010a, 2.

[7] Einstein to Paul Ehrenfest, 16 September 1925, as quoted in van Dongen 2007, 117.

When Einstein changed his recollections, he presented different histories of himself. This was not without programmatic intent, however. The 1918 statement on the pertinence of 'facts' was written in a personal letter to a friend, Michele Besso, but the 1949 recollection was offered quite publically: it was included in the *Autobiographical Notes* that were part of the highly visible volume "Albert Einstein: Philosopher-Scientist", dedicated to Einstein's science and philosophy. These *Notes* appear catered to promote Einstein's then long-standing research program in field theory unification for a well-chosen audience. The volume further contained a number of essays on physics and philosophy by prominent contemporaries to which Einstein wrote detailed replies. Its expected stature may be illustrated by the alternative two titles that the book's editor, Paul A. Schilpp, had contemplated: "(1) 'The Scientific Battle of The Twentieth Century,' and (2) 'The Future of Physics'" (Schilpp [1949], xvi). Clearly, Einstein, his co-contributors and their prospective readers were already well aware of his own status as opinion leader and iconic exemplar—in other words, of the prominent role that Einstein's biography played in the shared sense of what it meant to be a 'physicist', and a 'theoretical physicist' in particular.

The history of the young, experientially directed Einstein had become a familiar trope in epistemological discussions by the late 1920s. The logical empiricists, for example, had modeled their position that theories can be neatly divided in empirical propositions and coordinative definitions on Einstein's treatment of observation in relativity (Howard 1994). Likewise, experimental physicist Robert A. Millikan venerated Einstein on his seventieth birthday for having made "modern science essentially empirical", and he singled out the 1905 version of relativity which, according to Millikan, was supposedly created out of the notorious Michelson-Morley experiment on the (non-)existence of the ether (Millikan 1949). The older Einstein, however, actually distanced himself from logical empiricist philosophy and denied any direct role of the Michelson-Morley experiment in his creative process.[9] There are good reasons to doubt the older Einstein's recollections: for one, he had no problem attributing the Michelson-Morley experiment a prominent role in his earlier accounts of the events of 1905 (van Dongen 2009). Furthermore, the formulation of general relativity in 1915 certainly did depend on 'facts'.

Einstein's downplaying of the creative merit of experience went hand in hand with an increased emphasis on trusting "faith in […] logical simplicity" of the laws as guide.[10] This meant that researchers should aim for the mathematical unification of theories: they should aspire to bring together under one all-encompassing mathematical construction all of the natural forces (at the time, that meant the forces of gravity and electromagnetism), together with space, time and matter. As Einstein's research moved in the direction of this project, he consciously or unconsciously adjusted his recollections. By all appearances, he did so, at least in part, in order to justify his choices and to influence his field.

---

[8] Einstein to Michele Besso, 28 August 1918 (Speziali 1979, 81).

[9] Holton 1968, 1969; for a nuanced treatment of Einstein's relation to realism and empiricism, see also Howard 1993.

[10] Einstein to David Bohm, 24 November 1954, as quoted in van Dongen 2010a, 181-182; for more on Einstein and 'simplicity', see also Howard 1998.

The theory of quantum mechanics had been created with much direct input from experiment, unlike Einstein's efforts in unified field theory. He famously criticized quantum mechanics, initially for being inconsistent, and later for its alleged 'incompleteness' when considered from a realist perspective. Usually this aspect and the theories' probabilistic nature are cited as indications of a principally philosophical disagreement with the quantum theory. Yet, there is a complementary perspective which provides a more comprehensive view of Einstein's position: when assessing his positions on the quantum theory, one should also consider his positive program in field theory and indeed his ideas on how to properly do theoretical physics. The incompleteness of statistical descriptions was to be overcome by finding the proper equations for the "total field", i.e., logically simple and unified field equations and their ideal particle solutions (Einstein [1949], 81). As Einstein put it, the proper laws can foremost be found by "looking for the mathematically simplest concepts and the link between them. […] The creative principle resides in mathematics"; there "lies the theorist's hope of grasping the real in all its depth" (Einstein [1933], 300-301). Another way that Einstein expressed the same maxim was to state that one should strive after formulations of the laws that displayed "inner perfection" or "naturalness" (Einstein [1949], 23); in the end, only familiar classical field theories could possibly live up to his intuitions regarding 'naturalness'. These kinds of theories also safeguarded his own sense of physical causality and locality. In this sense, then, his familiar criticisms regarding the problematic nature of quantum mechanics should also be seen as an advertisement for his own efforts in field theory (see e.g. Einstein 1936).

Einstein judged quantum mechanics to be not only philosophically questionable but also mathematically unnatural. Werner Heisenberg's matrix formulation of the quantum theory, for example, was initially qualified as "a true witches' multiplication table."[11] Theory criticism was accompanied by implicit value judgments: the *Autobiographical Notes* assure us that "the faith" that experiential "facts by themselves can and should yield scientific knowledge" is based on a "prejudice" (Einstein 1949, 49). On another occasion, Einstein qualified the motivation for changing theories because of a conflict with experience "trivial, imposed from without"; unification and simplification of premises provided a "more subtle motive" (Einstein 1950, 13, 16). If the unification program fails, he confided to a young mathematician, Frenchman André Lichnerowicz, then we can only hope to understand things "superficially."[12] Einstein wrote another colleague that mere "humans are usually deaf to the strongest arguments, while they are constantly inclined to overestimate the accuracy of measurement."[13] The message is clear: the theorist who occupies himself with unified field theory is *virtuous*, as he wants more than superficial knowledge: Einstein insisted that a "true theorist" believes that comprehension is "built on premises of great simplicity" (Einstein 1950, 13).

---

[11] Einstein to Michele Besso, 25 December 1925 (Speziali 1979, 215-216).

[12] Einstein to André Lichnerowicz, January 1954, as quoted in van Dongen 2010a, 5.

[13] Einstein to Max Born, 12 May 1952 (Born et al. 1971, 188-189).

Where did Einstein's emphasis on these virtues come from, and when exactly did they become so very prominent? And when did he begin altering his recollections of the road to relativity? This is really a chicken and egg problem, but one moment in Einstein's biography does stand out as of great importance for the change in his epistemological outlook: finding the final field equations of the general theory of relativity in November of 1915. In his search for 'generally covariant' and relativistic field equations for gravity, Einstein had balanced a mathematical and physical approach, which initially led to contradictions. In 1913, he preferred the answers that his physical demands were giving him and he in fact more or less gave up on the mathematical idea of general covariance, despite that this went against one of his initial key intuitions. Eventually he returned to his more balanced approach and the final theory was again generally covariant. Einstein first concluded that he had too easily given up on his mathematics, but in fact he had needed both physics and mathematics (Norton 1984, Janssen and Renn 2007; Renn and Sauer 2007).

As we already saw, he did not immediately change his epistemological outlook: those changes would grow on him. Much later, for instance in a prominent 1933 lecture on methodology in Oxford, his 1949 autobiography or in the 1954 letter to de Broglie, the experience was brought forward to bulwark a strongly worded and a much more one-sided methodological conviction:

> The gravitational equations could *only* be found by a purely formal principle (general covariance), that is, by trusting in the largest imaginable logical simplicity of the natural laws. […] This is how I became a fanatic believer in the method of 'logical simplicity'.[14]

The methodological recipe of unified field theories resonated with, and partly originated in the experience of general relativity.

It should be pointed out that striving for unity and simplicity was also present in Einstein's earlier work, and it is not hard to identify origins of these ideas in even his earliest youth: Einstein had been an avid reader of the popular science books of Ludwig Büchner and Aaron Bernstein and these emphasized the banishment of dualisms and a search for unity in concepts (Gregory 2000). Striving for a unified *Weltbild* was not unique to Einstein, as many of his generation (and many more since then) shared this desire: regardless of whether they had as avidly enjoyed Bernstein and Büchner or not, Einstein's German contemporaries shared a cultural heritage that included Fichte, Goethe's Faust and the philosophies of Von Humboldt and Schelling, who all emphasized a search for a unified world picture (Holton 1998). In an article eulogizing a close colleague, Max Planck, Einstein even called this search the scientist's "supreme task" (Einstein 1918, 246). What makes the later Einstein stand out from his generation and the generation that immediately followed him, however, is his increased emphasis on the near-exclusive creative merit of mathematics and his increased distancing from experience in favor of theory.

---

[14] Einstein to Louis de Broglie, 8 February 1954, as quoted in van Dongen 2010a, 2-3; emphasis as in original.

How were Einstein's ideas about the ideal theory and ideal theorist put to use concretely? His field theory unifications spanned a period of at least three decades. They included such abstract attempts as five dimensional generalizations of relativity in the hope to accommodate the electromagnetic force alongside the gravitational force (now known as 'Kaluza-Klein' theories; see van Dongen 2002, 2010), or efforts to attain the same by considering space-time universes that were torqued rather than relativistically curved; these theories were called 'teleparallel' (Sauer 2006). All these theories were to reproduce the intrinsically discrete nature of the world captured by quantum mechanics and the integer valuedness of the electron charge. Presumably, their equations would allow solutions that avoided the problematic infinities typical for the general theory of relativity (known as 'singularities'). Einstein hoped that such solutions could model elementary particles in a natural fashion. He celebrated newly proposed equations invariably as of "great logical simplicity",[15] but abandoned them again when they did not produce the desired particle solutions. The virtue of 'simplicity' (or, in Einstein's usage, its near synonyms 'unity' and 'naturalness') guided these efforts, as the following words on his attempts at a mathematical unification of electrons and protons through so-called 'semivectors' testify:

> I discovered [...] a new sort of field, which we called 'semivectors'. These semivectors, after vectors, are the simplest mathematical fields that are possible in a four dimensional continuum, and they describe, in a natural way, essential properties of electrical particles. […] The important point is that all these constructions, and the laws connecting them, can be arrived at by the principle of looking for the mathematically simplest concepts and the link between them (Einstein 1933, 301).

Unfortunately for Einstein, the semivector result alluded to here proved to be spurious as well, and he quickly dropped the concept (van Dongen 2004, 2010).

**Unified field theory and the morally virtuous theorist**

Recounted recollection is a way to fashion our own self-image and an image of ourselves to the outside world. As we saw, in his autobiography and elsewhere, Einstein presented an edited version of his own history. In these accounts theoretical virtues and methodological precepts were reflected in the choices that a epistemically virtuous theorist was supposed to make. Einstein's example can show us yet more, however. We will argue that it makes clear that the ideal theoretical physicist was at the same time both morally and epistemically to be emulated: the virtuous had at the same time a moral *and* an epistemic charge, while epistemology carried religious connotations for Einstein.

Einstein considered choosing a more complicated mathematical structure over a simpler alternative to be 'sinful': "sin remains sin, even if it is committed by otherwise ever so respectable men" (Einstein 1949, 77). Objectionably complex were those theories that lacked a sufficient degree of unification. As we already saw, Einstein held up his senior Berlin colleague Max Planck as another example of a virtuous theoretical physicist. Planck had famously introduced the energy quantum in 1900 and Einstein wrote a eulogy for his sixtieth birthday in 1918. He expressed that Planck exhibited "inexhaustible patience and perseverance" while he "devoted himself […] to the most general problems of our science, refusing to let himself be diverted to more grateful and more easily attained ends." In Einstein's eyes, Planck was a unifier, too, who might yet "succeed in uniting quantum theory with electrodynamics and mechanics in a single logical system" (Einstein 1918, 247-248). In the 1918 Planck eulogy, Einstein still awarded experience a creative role in "uniquely determin[ing] the theoretical system"; in his 1949 autobiography, however, Einstein claimed that Planck's early work on the quantum was "all the more remarkable because, at least in its first phase, it was not in any way influenced by any surprising discoveries of experimental nature" (Einstein 1949, 37). As anyone familiar with the history of the black body radiation problem (Kuhn 1978, Gearheart 2002) will agree, this is simply wrong, as Einstein likely would have known too. His words seem intended to substantiate a controversial characteristic of the unified field theory methodology with another example, aside from his own, rather than an attempt at rendering Planck's efforts accurately. Indeed, Einstein also produced histories of Johannes Kepler and James Clerk Maxwell to convey similar messages (Einstein 1930a, 1931).

---
[15] Einstein to Maurice Solovine, 23 December 1938 (Einstein 1956, 76).

The 1918 Planck eulogy stands out for another remarkable characteristic. Its evocative language and imagery, which addressed the practice of science in an at times exalted fashion, is easily compared to religiously inspired moral text. In the "temple of science", as Einstein chose his words, only true scientists would remain if "an angel of the Lord" were to drive out all the Philistines. Those who remain might fulfill that "supreme task […] to arrive at those universal elementary laws from which the cosmos can be built up by pure deduction" (Einstein 1918, 244, 247). Here, religious metaphor emphasized the virtuous nature required to practice science properly: only those who aim at the universal—at "unity in the foundations" (Einstein 1936)—are truly virtuous. Belief in such methodology bordered on religious belief: in 1950, Einstein wrote that the idea "that the totality of all sensory experience can be 'comprehended' on the basis of a conceptual system built on premises of great simplicity" was indeed "a miracle creed".[16] (Einstein 1950, 13; on this article, see particularly also Sauer 2015).

Einstein also used exalted language in the Autobiographical Notes: he spoke of a "paradise" beyond the "merely-personal", the "contemplation" of which "beckoned like a liberation" in his early youth (Einstein 1949, 5). According to Lorraine Daston (2008), such abnegations of the personal are to be seen as an expression of the rise of objectivity during the same period. We should be aware, however, that to Einstein, the romantic genius is more of an ideal type than the self-denying record keeper. Einstein's paradise beyond the personal is not found in unfiltered maps of data points and immediate representations of observations, but in the unified universes captured by grand and speculative thought. As he wrote in his article on *Religion and science*: "individual existence [is] a sort of prison and one wants to experience the universe as a single significant whole." Not the nitty-gritty of experimental detail, but the leap of abstract and intuitive creativity motivates and liberates. Indeed, those "geniuses" that may attain such elevation possess a "cosmic religious feeling" (Einstein [1930b], 41). In this light, Einstein's scientific flight from the confining personal and "primitive feelings" (Einstein 1949, 5) was intended to move him beyond the partial and particular of incomplete descriptions: he wished to attain the impressive breadth and deductive certainty of all-encompassing and fully unified theories.

---
[16] As in Einstein 1950, 13; on this article, see particularly also Sauer 2015.

In the history of objectivity, Arthur Schopenhauer figures prominently. Daston and Galison believe that his instruction to suppress the 'Will', as a precondition to enable introspective reflection towards the apprehension of the principles of the world, is related to the objective scientist's self-restraint—his philosophy was a contemporary expression of the same sentiment (Daston and Galison 2007, 203-204). The objective persona, Daston (in a separate publication) further finds, eschewed the personal while immersed in solitudinous thought—yet, at the same time, he was part of a timeless community of scholars, who were imagined as equally "thinking selves dedicated to fathoming nature" (Daston 2008, 17). These scholars, then, would produce philosopher Thomas Nagel's 'view from nowhere': the same timeless and invariant facts and structures were to be unearthed by them, independent of any personal dimensions. In Daston's account, here is where Einstein fits the story of objectivity: his "paradise beyond the personal" could be attained by aiming at invariance and symmetry, which would be the same to all but only revealed to a few.

Einstein's unified field theory agenda and refashioned ideal theorist do resonate with this aspect of the history of objectivity: both are equally aimed at the universal. Indeed, Einstein approvingly identified Schopenhauer as someone who had understood that the "cosmic religious feeling" originates in the desire to "experience the universe as a single significant whole": in the "sublimity and marvelous order which reveal themselves in nature and in the world of thought" does one, à la Schopenhauer, find a way to escape "the futility of human desires" (Einstein [1930b], 41). Einstein believed, "with Schopenhauer", that the ideal scholar, "a finely tempered nature", should wish to "escape from personal life into the world of objective perception and thought." That perception, however, is not the unmediated communication of direct observation, but the contemplative view on the impersonal and logically simple, "the restful contours apparently built for eternity" (Einstein [1918], 245, 251).

Another philosopher, at least as much read and revered by Einstein, captured a similar sentiment, even if he wrote and thought in an entirely different era: Baruch Spinoza. In the Weimar years Spinoza's philosophy went through a remarkable revival among German Jews as he was widely celebrated as a Jewish hero. His biography offered Jews the possibility to celebrate increasingly embattled liberal values and to still believe in assimilation, while affirming their ethnic authenticity as was particular to the Jewish renaissance of the Weimar years (Wertheim 2011). Einstein had read Spinoza throughout his life, beginning right after finishing his college education together with his friends of the mock 'Akademie Olympia', and on many occasions thereafter, even visiting Spinoza's small house in Rijnsburg near Leiden in 1920 (Stachel et al. 1989; Buchwald et al. 2006, 604); historian of science Max Jammer has actually identified Spinoza as Einstein's "favorite philosopher" (Jammer 1999, 144).

In Spinoza, rational contemplation of the world, moral virtue and a unified understanding of God and nature meet: Einstein's approval of his philosophy is an expression of what it meant to him to be both a virtuous scientist and a morally just person. Epistemic and moral virtue overlap here in a most direct manner. Spinoza, in his *Ethics*, argued (with geometric precision, that is, in a logical style modeled on Euclid's geometry) that man's moral well-being lies not in a life spent on pursuing the lower passions or the contemplation of the superstitions known as religion, but in a life spent on the rational pursuit of knowledge, particularly knowledge of the eternal laws of nature. Nature, in Spinoza's view, a view which Einstein warmly agreed with, was synonymous to an impersonal God whose existence was infinite and deterministic. According to Spinoza, our virtue, happiness and even salvation lies in understanding the universe: the more we know of the true ideas of God, the freer we are, since we will deal with life's disturbances with greater equanimity (Nadler 2001). Spinoza, too, saw in rational contemplation of unified eternity a flight from the "merely personal." Einstein agreed with Spinoza's positions, identified with Spinoza as a fellow Jew, and most famously followed him in his understanding of God: "I believe in Spinoza's God who reveals himself in the orderly harmony of what exists, not in a God who concerns himself with fates and actions of human beings."[17]

Einstein held that our moral and esthetic judgment are "tributary forms in helping the reasoning faculty toward its highest achievements"—one is right in "speaking of the moral foundations of science", even if the reverse, to reduce ethics to formulae, "must fail."[18] Science cannot give us a moral road map, but it provides a way to elevate man: "one does people the best service by giving them some elevating work to do and thus indirectly elevating them", Einstein wrote in a essayistic note entitled "Good and evil." He specified that such work could be in the arts or sciences, and in the latter case added that "to be sure, it is not the fruits of scientific research that elevate a man and enrich his nature, but the urge to understand" (Einstein [1934], 13). This reminds us of both Spinoza's *Ethics* and Einstein's unified field theory efforts—not just because Einstein asked us to leave out of consideration whether one's contemplations bear fruit, but primarily because of the insistence on "understanding." For, what did he believe "understanding" entailed? In Einstein's physics, to understand meant to have a 'simple' and unified description: "[We mean] by simplicity [...], that the system contains as few independent assumptions or axioms as possible; for the totality of logically independent axioms stands for the 'un-understood' remainder." In the same 1931 essay, he further made clear that his own efforts "first and foremost concern the logical unity in physics." Indeed, because "one wants to *understand* the existing, real world."[19] So Einstein's ideal theorist combines both epistemic and moral virtue in his attempt to unify: unified field theory science is virtuous science à la Spinoza, as it aims to elevate by aspiring to understand the universe. At the same time, to adhere to quantum mechanics was not virtuous, as it did not deliver 'understanding'. Einstein confided to a young fellow critic of the quantum theory, David Bohm, that "I believe that [the unified] laws are logically simple and that the faith in this logical simplicity is our best guide, in the sense that it suffices to start from relatively little empirical knowledge. If nature is not arranged in a way corresponding to this belief, then there is no hope at all to arrive at a deeper *understanding*."[20] As we saw, without such understanding, man will remain bound to the confines of the superficial and personal.

---

[17] Einstein to Rabbi Herbert S. Goldstein, April 1929, on p. 49 in Jammer 1999.

[18] Einstein, 1930, interview cited in Jammer 1999, p. 69.

[19] Einstein in unpublished essay document, dated 1931. Albert Einstein Archives, Jerusalem, entry no. 2-110; emphasis as in original.

[20] Einstein to David Bohm, 24 November 1954, on pp. 181-182 in van Dongen 2010a.

The close relation between religious sensibility, and moral and epistemic virtue was expressed well by Spinoza in Einstein's assessment: "those individuals to whom we owe the great creative achievements of science were all of them imbued with the truly religious conviction that this universe of ours is something perfect and susceptible to the striving for knowledge. If this conviction had not been a strongly emotional one and if those searching for knowledge had not been inspired by Spinoza's *Amor Dei Intellectualis*, they would hardly have been capable of that untiring devotion which alone enables man to attain his greatest achievements" (Einstein [1948], 56-57). Einstein insisted on the devotion of science's practitioners: He held that "the cosmic religious feeling" was experienced when the "futility of human desires and the sublimity and marvelous order which reveals itself both in nature and in the world of thought" is recognized (Einstein [1930b], p. 41). Just as Spinoza, Einstein saw personal transcendence to take place in intellectual study of nature, while thought and nature overlapped: "I have found no better expression for 'religious' than confidence in the rational nature of reality as it is accessible to human reason. Wherever this feeling is absent, science degenerates into uninspired empiricism."[21] So, mathematical unification stands to quantum mechanical empiricism as 'religious' stands to 'uninspired'. Einstein considered himself religious in a way that resonated with both Spinoza and his unification program: "my feeling is insofar religious as I am imbued with the consciousness of the insufficiency of the human mind to understand deeply the harmony of the Universe which we try to formulate as 'laws of nature'"; to attain such a "consciousness", "humility" was required.[22] Only when the "futility of human desires" is acknowledged will we "experience the universe as a single, significant whole"—then, moral, epistemic and personal fulfilment may be attained.

**Conclusion**

Remarkably, Spinoza is not mentioned in Einstein's *Autobiographical Notes*. Still, the ideas expressed there exhibit the same interdependence of striving for unification in physics, moral and epistemic virtue, and personal deliverance, while moral and epistemic qualities particularly met in the scientist's religious gaze. At the same time, Einstein rewrote his own history and crafted his own version of his persona to justify and promote his unification attempts: theoretical and personal virtues here mirrored one another.

It need not surprise us, then, that scholars involved in quantum mechanics and its successors, when rejecting Einstein's unified field theories, not only rejected the science but also his ideas of what a theorist was supposed to be, along with his own embodiment of those ideas. Robert Oppenheimer, for instance, deemed the older Einstein as someone who had "lost contact with the profession of physics" and as "completely cuckoo." Einstein would have been "wasting his time", Oppenheimer found (Schweber 2008, 276, 279; Bird and Sherwin 2006, 64). At stake in these judgements was what it meant to be a theorist, and what a theorist's relation to the empirical is. Clearly, it is indeed in efforts to understand the formation of disciplinary identity that personae and epistemic virtues are necessary as historiographical tools, just as they are most useful in understanding intra-disciplinary debates over future directions, as Oppenheimer's responses illustrate.

---

[21] Einstein to Maurice Solovine, 1 January 1951 (Einstein 1956, 102-104).

[22] Einstein to B[?] F[?], 17 December 1952, quoted as in Jammer 1999, 122; see also Holton 2005.

What we particularly wished to argue here is that concerns over what it meant to be a theorist were reflected in the theories constructed: a 'superficial' scholar, in Einstein's judgement, aimed for the empirical and could be content with a theory that was only concerned with probabilities and not full reality. Humble, devoted and "true" theorists, on the other hand, produced theories that were unified and mathematically natural. They were not distracted by "the immediately given."[23] The example of Einstein shows that tracking idealized 'virtuous theorists' helps to identify histories of theoretical virtues, and how they operated in theory choice. We further saw that virtue is at the same time morally and epistemically charged.

It has been expressed by philosophers of science, in frustration with the contingencies emphasized by historiography, that Thomas Kuhn's account of the role of values in theory choice should be cleansed of words with moral connotations—that instead of 'virtues' and 'values' a morally neutral and less historically contingent term as epistemic 'conditions' should be preferred when describing theory choice.[24] Yet, when we break the connection between theory choice, epistemic virtues and personae, we risk to lose sight of the people in our histories, particularly as scholars express and argue their perspective through morally charged, idealized biography. Einstein's example, at least, makes clear that Kuhn's values in theory choice should be thoroughly historicized: only then can we unearth the interweaving of epistemic and moral virtues and their role played in assessing theory.

**Bibliography**

*Primary sources*

Born, Max, Hedwig Born and Albert Einstein. [1971] 2005. The Born-Einstein Letters. New York: Macmillan.

Buchwald, Diana, Tilman Sauer, Ze'ev Rosenkranz, József Illy, Virginia I. Holmes, Jeroen van Dongen, Daniel J. Kennefick and A.J. Kox. 2006. *The Collected Papers of Albert Einstein*, Volume 10: *The Berlin Years, Correspondence May-December 1920 and Supplementary Correspondence, 1909-1920*. Princeton: Princeton University Press.

Einstein, Albert. 1905. Zur Elektrodynamik bewegter Körper. *Annalen der Physik* 17: 891-920.

_____. [1918] 1994. Principles of Research. In *Ideas and Opinions*, 244-248. New York: The Modern Library.

_____. [1930a] 1994. Johannes Kepler. In *Ideas and Opinions*, 286-291. New York: The Modern Library.

---

[23] "Immediately given" is a term used by Einstein, see e.g. his letter to Moritz Schlick, 21 May 1917 (Schulmann et al. 1998, 456-457).

[24] This point was debated at the excellent conference 'Theoretical Virtues in Theory Choice', 12-14 July 2012, at the Zukunftskolleg of the University of Konstanz.

_____. [1930b] 1994. Religion and Science. In *Ideas and Opinions*, 39-43. New York: The Modern Library.

_____. [1931] 1994. Maxwell's Influence on the Evolution of the Idea of Physical Reality. In *Ideas and Opinions*, 291-295. New York: The Modern Library.

_____. [1933] 1994. On the Method of Theoretical Physics. In *Ideas and Opinions*, 296-303. New York: The Modern Library.

_____. [1934] 1994. Good and Evil. In *Ideas and Opinions*, 13. New York: The Modern Library.

_____. 1936. Physik und Realität. *Journal of the Franklin Institute* 221: 313-347.

_____. [1948] 1994. Religion and Science: Irreconcilable? In *Ideas and Opinions*, 53-57. New York: The Modern Library.

_____. [1949] 1997. Autobiographisches [and translation: Autobiographical Notes]. In *Albert Einstein: Philosopher-Scientist*, ed. Paul A. Schilpp, 1-94. La Salle, IL: Open Court.

_____. 1950. On the Generalized Theory of Gravitation. *Scientific American* 182: 13-17.

_____. 1956. *Lettres à Maurice Solovine*. Paris: Gauthier-Villars.

Heisenberg, Werner. [1974] 1989. Encounters and Conversations with Albert Einstein. In *Encounters with Einstein and Other Essays on People, Places and Particles*, 107-122. Princeton: Princeton University Press.

Millikan, Robert A. 1949. Albert Einstein on His Seventieth Birthday. *Reviews of Modern Physics* 21: 343-345.

Schilpp, Paul A. [1949] 1997. Preface. In *Albert Einstein: Philosopher-Scientist*, ed. Paul A. Schilpp, xiii-xvi. La Salle, IL: Open Court.

Schulmann, Robert, A.J. Kox, Michel Janssen and József Illy. 1998. *The Collected Papers of Albert Einstein*, Volume 8: *The Berlin Years, Correspondence, 1914-1917*. Princeton: Princeton University Press.

Speziali, Pierre, ed. 1979. *Albert Einstein, Michele Besso. Correspondance 1903-1955*. Paris: Hermann.

Stachel, John, David C. Cassidy, Jürgen Renn, Robert Schulmann, Don Howard and A.J. Kox. 1989. *The Collected Papers of Albert Einstein*, Volume 2: *The Swiss Years, Writings: 1900-1909*. Princeton: Princeton University Press.

***Secondary sources***

Bird, Kai and Martin J. Sherwin. 2006. *American Prometheus: The Triumph and Tragedy of J. Robert Oppenheimer*. New York: Vintage.

Daston, Lorraine. 2008. A Short History of Einstein's Paradise Beyond the Personal. In *Einstein for the 21st Century: His Legacy in Science, Art, and Modern Culture*, ed. Peter Galison, Gerald Holton and Silvan Schweber, 15-26. Princeton: Princeton University Press.

_____ and Peter Galison. 2007. *Objectivity*. New York: Zone Books.

_____ and Otto Sibum. 2003. Introduction: Scientific Personae and Their Histories. *Science in Context* 16: 1-8.

Gearheart, Clayton. 2002. Planck, the Quantum, and the Historians. *Physics in Perspective* 4: 170-215.

Gregory, Frederick. 2000. The Mysteries and Wonders of Natural Science: Aaron Bernstein's *Naturwissenschaftliche Volksbücher* and the Adolescent Einstein. In *Einstein: The Formative Years, 1879-1909*, eds. Don Howard and John Stachel, 23-42. Boston: Birkhäuser.

Holton, Gerald. 1968. Mach, Einstein and the Search for Reality. *Daedalus* 97: 636-673.

_____. 1969. Einstein, Michelson, and the 'Crucial' Experiment. *Isis* 60: 133-197.

_____. 1998. Thematic Presuppositions and the Direction of Scientific Advance. In *The Advancement of Science, and its Burdens*, 3-27. Cambridge, MA: Harvard University Press.

_____. 2005. Einstein's Third Paradise. In *Victory and Vexation in Science. Einstein, Bohr, Heisenberg and Others*, 3-15. Cambridge, MA: Harvard University Press.

Howard, Don. 1993. Was Einstein Really a Realist? *Perspectives on Science* 1: 204–51.

_____. 1994. Einstein, Kant, and the Origins of Logical Empiricism. In *Language, Logic, and the Structure of Scientific Theories*, ed. Wesley Salmon and Gereon Wolters, 45-105. Pittsburgh: University of Pittsburgh Press & Konstanz: Universitätsverlag.

_____. 1998. Astride the Divided Line: Platonism, Empiricism, and Einstein's Epistemological Opportunism. In *Idealization in Contemporary Physics*, ed. Niall Shanks, 143–63. Amsterdam and Atlanta: Rodopi.

Jammer, Max. 1999. *Einstein and Religion.* Princeton: Princeton University Press.

Janssen, Michel and Jürgen Renn. 2007. Untying the Knot: How Einstein Found His Way Back to Field Equations Discarded in the Zurich Notebook. In *The Genesis of General Relativity*, Volume 2: *Einstein's Zurich Notebook: Commentary and Essays*, ed. M. Janssen, J. Norton, J. Renn, T. Sauer and J. Stachel, 839-925. Dordrecht: Springer.

Jungnickel, Christa, and Russell McCormmach. 1986. *Intellectual Mastery of Nature: Theoretical Physics from Ohm to Einstein*, 2 vols. Chicago: University of Chicago Press.

Kuhn, Thomas S. 1977. Objectivity, Value Judgment and Theory Choice. In *The Essential Tension*, 320-339. Chicago: University of Chicago Press.

_____. 1978. *Black Body Theory and the Quantum Discontinuity, 1894-1912*. Oxford: Clarendon Press.

Nadler, Steven. 2001. Baruch Spinoza. Stanford Encyclopedia of Philosophy. https://plato.stanford.edu/entries/spinoza/. Accessed 23 December 2016.

Norton, John. 1984. How Einstein Found His Field Equations, 1912-1915. *Historical Studies in the Physical Sciences* 31: 231-316.

Renn, Jürgen and Tilman Sauer. 2007. Pathways Out of Classical Physics: Einstein's Double Strategy in His Search for the Gravitational Field Equation. In *The Genesis of General*

*Relativity*, Volume 1: *Einstein's Zurich Notebook: Introduction and Source*, ed. M. Janssen, J. Norton, J. Renn, T. Sauer and J. Stachel, 113-312. Dordrecht: Springer.

Sauer, Tilman. 2006. Field Equations in Teleparallel Space-time: Einstein's Fernparallelismus Approach Toward Unified Field Theory. *Historia Mathematica* 33: 399-439.

_____. 2015. Einstein's Unified Field Theory Program. In *The Cambridge Companion to Einstein*, ed. Michel Janssen and Christoph Lehner, 281-305. Cambridge: Cambridge University Press.

Schweber, Silvan S. 2008. *Einstein and Oppenheimer: The Meaning of Genius*. Cambridge, MA: Harvard University Press.

Tai, Chaokang. 2017. Left Radicalism and the Milky Way: Connecting the Scientific and Socialist Virtues of Anton Pannekoek. *Historical Studies in the Natural Sciences* 47: 200-254.

_____ and Jeroen van Dongen. 2016. Anton Pannekoek's Epistemic Virtues in Astronomy and Socialism: Personae and the Practice of Science. *Low Countries Historical Review* 131: 55–70.

Van Dongen, Jeroen. 2002. Einstein and the Kaluza-Klein Particle. *Studies in History and Philosophy of Modern Physics* 33: 185-210.

_____. 2004. Einstein's Methodology, Semivectors and the Unification of Electrons and Protons. *Archive for History of Exact Sciences* 58: 219-254.

_____. 2007. Emil Rupp, Albert Einstein, and the Canal Ray Experiments on Wave-Particle Duality: Scientific Fraud and Theoretical Bias. *Historical Studies in the Physical and Biological Sciences* 37 Suppl.: 73-120.

_____. 2009. On the Role of the Michelson-Morley Experiment: Einstein in Chicago. *Archive for the History of Exact Sciences* 63: 655-663.

_____. 2010a. *Einstein's Unification*. Cambridge: Cambridge University Press.

_____. 2010b. On Einstein's Opponents, and Other Crackpots. *Studies in History and Philosophy of Modern Physics* 41: 78-80.

Wazeck, Milena. 2009. *Einsteins Gegner. Die öffentliche Kontroverse um die Relativitätstheorie in den 1920er Jahren*. Frankfurt a.M.: Campus.

Wertheim, David J. 2011. *Salvation through Spinoza: A Study of Jewish Culture in Weimar Germany*, Leiden: Brill.